\documentclass{jpp}

\newcommand{\apj}{\mbox{\it Astrophysical Journal}}

\newcommand{\pre}{\mbox{\it Phys. Rev. E}}

\usepackage{color}
\usepackage{graphicx}
\usepackage{natbib}

\ifCUPmtlplainloaded \else
  \checkfont{eurm10}
  \iffontfound
    \IfFileExists{upmath.sty}
      {\typeout{^^JFound AMS Euler Roman fonts on the system,
                   using the 'upmath' package.^^J}%
       \usepackage{upmath}}
      {\typeout{^^JFound AMS Euler Roman fonts on the system, but you
                   dont seem to have the}%
       \typeout{'upmath' package installed. JPP.cls can take advantage
                 of these fonts, if you use 'upmath' package.^^J}%
      }
  \else
  \fi
\fi

\ifCUPmtlplainloaded \else
  \checkfont{msam10}
  \iffontfound
    \IfFileExists{amssymb.sty}
      {\typeout{^^JFound AMS Symbol fonts on the system, using the
                'amssymb' package.^^J}%
       \usepackage{amssymb}%

      }{}
  \fi
\fi

\ifCUPmtlplainloaded \else
  \IfFileExists{amsbsy.sty}
    {\typeout{^^JFound the 'amsbsy' package on the system, using it.^^J}%
     \usepackage{amsbsy}}
    {}
\fi

\newsavebox{\astrutbox}
\sbox{\astrutbox}{\rule[-5pt]{0pt}{20pt}}

\newcommand{\rev}[1]{{\color{red}#1}}

\title[Density jump as a function of magnetic field for switch-on collisionless shocks]{Density jump as a function of magnetic field for switch-on collisionless shocks in pair plasmas}

\author[A. Bret and R. Narayan]%
{Antoine Bret$^{1,2}$, Ramesh Narayan$^{3,4}$%
  \thanks{Email address for correspondence: antoineclaude.bret@uclm.es}
}

\affiliation{$^1$ETSI Industriales, Universidad de Castilla-La Mancha, 13071 Ciudad Real, Spain\\[\affilskip]
$^2$Instituto de Investigaciones Energ\'{e}ticas y Aplicaciones Industriales, Campus Universitario de Ciudad Real, 13071 Ciudad Real, Spain\\[\affilskip]
$^3$Harvard-Smithsonian Center for Astrophysics, Harvard University, 60 Garden St., Cambridge, MA 02138 USA\\
$^4$Black Hole Initiative at Harvard University, 20 Garden Street, Cambridge, MA 02138, USA
}

\date{?; revised ?; accepted ?. - To be entered by editorial office}

\begin{document}

\maketitle

\begin{abstract}
The properties of collisionless shocks, like the density jump, are usually derived from magnetohydrodynamics (MHD), where isotropic pressures are assumed. Yet, in a collisionless plasma, an external magnetic field can sustain a stable anisotropy. In \cite{BretJPP2018}, we devised a model for the kinetic history of the plasma through the shock front, allowing to self-consistently compute the downstream anisotropy, hence the density jump, in terms of the upstream parameters. This model dealt with the case of a parallel shock, where the magnetic field is normal to the front both in the upstream and the downstream. Yet, MHD also allows for shock solutions, the so-called switch-on solutions, where the field is normal to the front only in the upstream. This article consists in applying our model to these switch-on shocks. While MHD offers only 1 switch-on solution within a limited range of Alfv\'{e}n Mach numbers, our model offers 2 kinds of solutions within a slightly different range of Alfv\'{e}n Mach numbers. These 2 solutions are most likely the outcome of the intermediate and fast MHD shocks under our model. While the intermediate and fast shocks merge in MHD for the parallel case, they do not within our model. For simplicity, the formalism is restricted to non-relativistic shocks in pair plasmas where the upstream is cold.
\end{abstract}

\maketitle

\section{Introduction}
Shock waves are fundamental processes in plasmas which are usually studied within the context of magnetohydrodynamics (MHD). As an extension of fluid dynamics to plasmas, MHD entails the same assumption of small mean-free-path (see for example \cite{gurnett2005} \S 5.4.4, \cite{Goedbloed2010} chapters 2 \& 3, or  \cite{TB2017} \S 13.2). When fulfilled, collisions ensure that the pressure is isotropic both in the upstream and downstream, which simplifies the conservation equations.

In collisionless shock, where the mean-free-path is larger than the size of the system, the isotropy assumption may not be fulfilled, possibly resulting in a departure from the MHD predicted behavior. Such is especially the case in the presence of an external magnetic field which can stabilize a temperature anisotropy, as has been observed in the solar wind \citep{BalePRL2009,MarucaPRL2011,SchlickeiserPRL2011} and is projected to be studied in the laboratory \citep{Carter2015APS}.

Some authors worked out the MHD conservation equations in the case of anisotropic pressure, and studied the consequences on the shock properties \citep{Erkaev2000,Double2004,Gerbig2011}. Yet, in these works, while the upstream is assumed isotropic, the downstream degree of anisotropy is left as a free parameter.

Recently, a self-contained theory of magnetized collisionless shocks has been developed. By making some assumptions on the kinetic history of the plasma as it crosses the front, we could compute the downstream degree of anisotropy, for the parallel and the perpendicular cases, in terms of the magnetic field strength \citep{BretJPP2018,BretPoP2019,BretLPB2020}.

Noteworthily, the theory for parallel shocks described in \cite{BretJPP2018}  has been successfully tested against Particle-In-Cell (PIC) simulations in \cite{Haggerty2022}.

In MHD, several shock solutions exist when the upstream magnetic field is aligned with the flow. The most common solution is the one where the downstream field is also aligned with the flow. This is the fully parallel case, where the fluid and the field are decoupled \citep{Lichnerowicz1976,Majorana1987}. Yet, still for the case where the upstream field is parallel to the flow, MHD offers a second option: the switch-on shocks \citep{fitzpatrick,Kulsrud2005,Goedbloed2010}. In such shocks, while the magnetic field does not have any components along the shock front in the upstream, it has one in the downstream. Indeed, the MHD conservation equations only enforce the continuity of the field component perpendicular to the front, not the continuity of the normal component. Therefore, they allow for solutions, the switch-on solutions, where the upstream field is normal to the front while the downstream field is not.

The theory developed in \cite{BretJPP2018} was the collisionless version of the fully parallel MHD case. The present article deals with the collisionless version of the MHD switch-on shocks.

As in \cite{BretJPP2018}, we consider, for simplicity, pair plasmas for which both species have the same perpendicular and parallel temperatures to the field. In Section \ref{sec:MHD}, we remind the MHD results for switch-on shocks. In Section \ref{sec:Method}, we explain the method used. It significantly differs from \cite{BretJPP2018} since we need to account for an oblique downstream field. In addition, MHD results suggest the obliquity of the downstream field, labelled $\theta_2$ in the sequel, can be as high as $0.56 \frac{\pi}{2}$ (see Fig. \ref{fig:theta2switchon}-left). We cannot therefore work out a theory restricted to $\theta_2 = \varepsilon$, with $0 < \varepsilon \ll 1$. Then, in Sections \ref{sec:S1} \& \ref{sec:S2}, we explain the solutions found for switch-on shocks within our model.

\begin{figure}
\begin{center}
 \includegraphics[width=\textwidth]{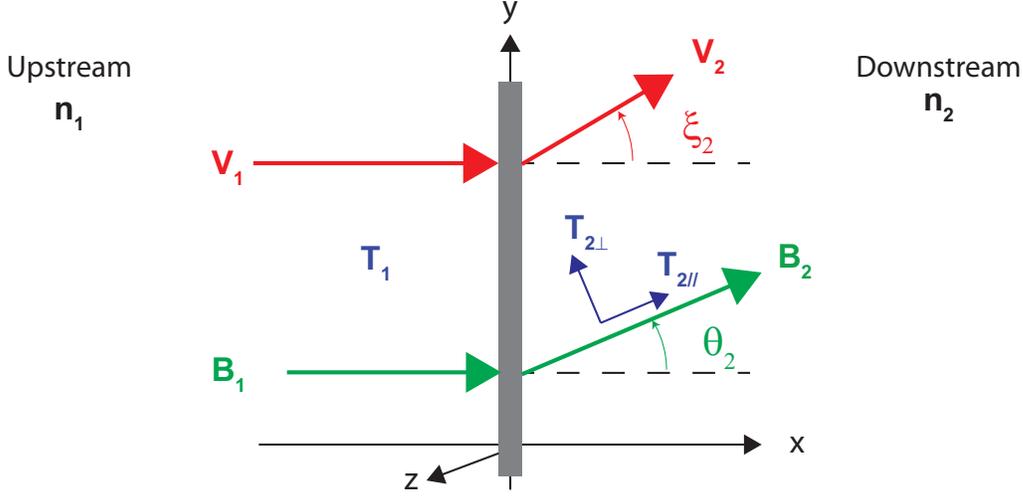}
\end{center}
\caption{System considered. The upstream has density $n_1$ and isotropic temperature $T_1$. Both the upstream field $\mathbf{B}_1$ and velocity $\mathbf{v}_1$ are normal to the front. The downstream has density $n_2$ and  temperatures $T_{2\parallel}, T_{2\perp}$, parallel and perpendicular to the downstream field. The downstream field $\mathbf{B}_2$ and velocity $\mathbf{v}_2$ make an angle $\theta_2$ and $\xi_2$ respectively with the front normal. The parallel and perpendicular directions are therefore defined with respect to the local magnetic field.}\label{fig:system}
\end{figure}

\section{MHD results}\label{sec:MHD}
The system considered is sketched on Figure \ref{fig:system}. The upstream field $\mathbf{B}_1$ and velocity $\mathbf{v}_1$ are normal to the front, but the downstream ones $\mathbf{B}_2$ and $\mathbf{v}_2$ are not. They make an angle $\theta_2$ and $\xi_2$ with the shock normal and by default, $\theta_2 \neq \xi_2$ (even though they will be found equal in the sequel).

We here briefly remind the MHD theory for switch-on shocks. Due to the complexity of the forthcoming calculations, we treat only the case of a sonic strong shock, namely upstream temperature $T_1=0$, or equivalently, upstream sonic Mach number $\mathcal{M}_{s1}=\infty$.

For isotropic pressures in the upstream and the downstream, and $\theta_1=\xi_1=0$, the MHD conservation equations for strong shock and an adiabatic index of $\gamma=5/3$ read (see for example \cite{Kulsrud2005}, p. 141),
\begin{eqnarray}
n_2 v_2 \cos \xi_2=&n_1 v_1, \label{eq:mhd1}  \\
B_2 \cos \theta_2=&B_1 ,       \label{eq:mhd2}  \\
B_2 v_2 \sin \theta_2 \cos \xi_2 - B_2 v_2 \cos \theta_2 \sin \xi_2=& 0,      \label{eq:mhd3}  \\
\frac{B_2^2 \sin ^2\theta_2}{8 \pi }+n_2 k_B T_2 + m n_2 v_2^2 \cos ^2\xi_2  =&  m n_1 v_1^2, \label{eq:mhd4}\\
 m n_2 v_2^2 \sin \xi_2 \cos \xi_2-\frac{B_2^2 \sin \theta_2 \cos \theta_2}{4 \pi }  =& 0, \label{eq:mhd5}\\
m n_2 v_2 \cos \xi_2 \left(\frac{5}{2}\frac{k_B T_2 }{m}+\frac{B_2^2 \sin ^2\theta_2}{4 \pi  m n_2}+\frac{v_2^2}{2}\right)
-\frac{B_2^2}{4 \pi }v_2 \sin \theta_2 \cos \theta_2 \sin \xi_2 =& \frac{1}{2} m n_1 v_1^3 , \label{eq:mhd6}
\end{eqnarray}
where $m$ is the mass of the particles and $k_B$ the Boltzmann constant.

Eq. (\ref{eq:mhd1}) stands for the conservation of mass. Eq. (\ref{eq:mhd2}) for the conservation of the magnetic field normal component.  Eq. (\ref{eq:mhd3}) for the vanishing of the $z$ component of the electric field. Eqs. (\ref{eq:mhd4},\ref{eq:mhd5}) come from the conservation of the momentum flux (see Appendix \ref{ap:cons}), and Eq. (\ref{eq:mhd6}) from the conservation of energy.

By eliminating $v_2$ and $B_2$ thanks to Eqs. (\ref{eq:mhd1},\ref{eq:mhd2}), and then eliminating $T_2$ thanks to Eq. (\ref{eq:mhd4}), the system is amenable to 3 equations,
\begin{eqnarray}
  \tan \theta_2-\tan \xi_2 &=& 0 ,   \label{eq:mhd3_1}\\
  \mathcal{M}_{A1}^2 \tan \xi_2-r \tan \theta_2 &=&  0 , \label{eq:mhd3_2} \\
 2 \mathcal{M}_{A1}^2 \left[(r-5) r+5-\sec ^2\xi_2\right]+r \tan \theta_2 \left(\tan \theta_2+4 \tan \xi_2\right)  &=&  0 \label{eq:mhd3_3},
\end{eqnarray}
in terms of the dimensionless density ratio $r$ and the Alfv\'{e}n Mach number $\mathcal{M}_{A1}$,
\begin{eqnarray}\label{eq:dimless}
  r &=& \frac{n_2}{n_1}, \\
  \mathcal{M}_{A1}^2 &=& \frac{m n_1 v_1^2}{B_1^2/4\pi}. \nonumber
\end{eqnarray}
The first equation imposes $\theta _2=\xi _2$. Replacing in the last 2 gives,
\begin{eqnarray}
  \tan \theta_2 \left(\mathcal{M}_{A1}^2-r\right) &=& 0, \label{eq:1MHD}  \\
  2 \mathcal{M}_{A1}^2 \left[(r-5) r+5-\sec ^2\theta_2\right]+5 r \tan ^2\theta_2 &=& 0 \label{eq:2MHD}.
\end{eqnarray}

\begin{figure}
  \centering
  \includegraphics[width=0.5\textwidth]{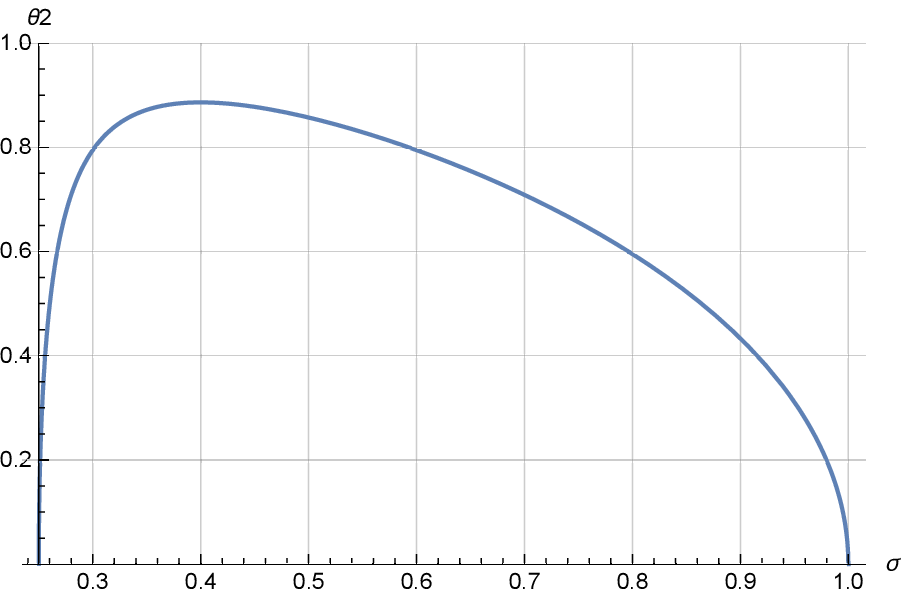}\includegraphics[width=0.5\textwidth]{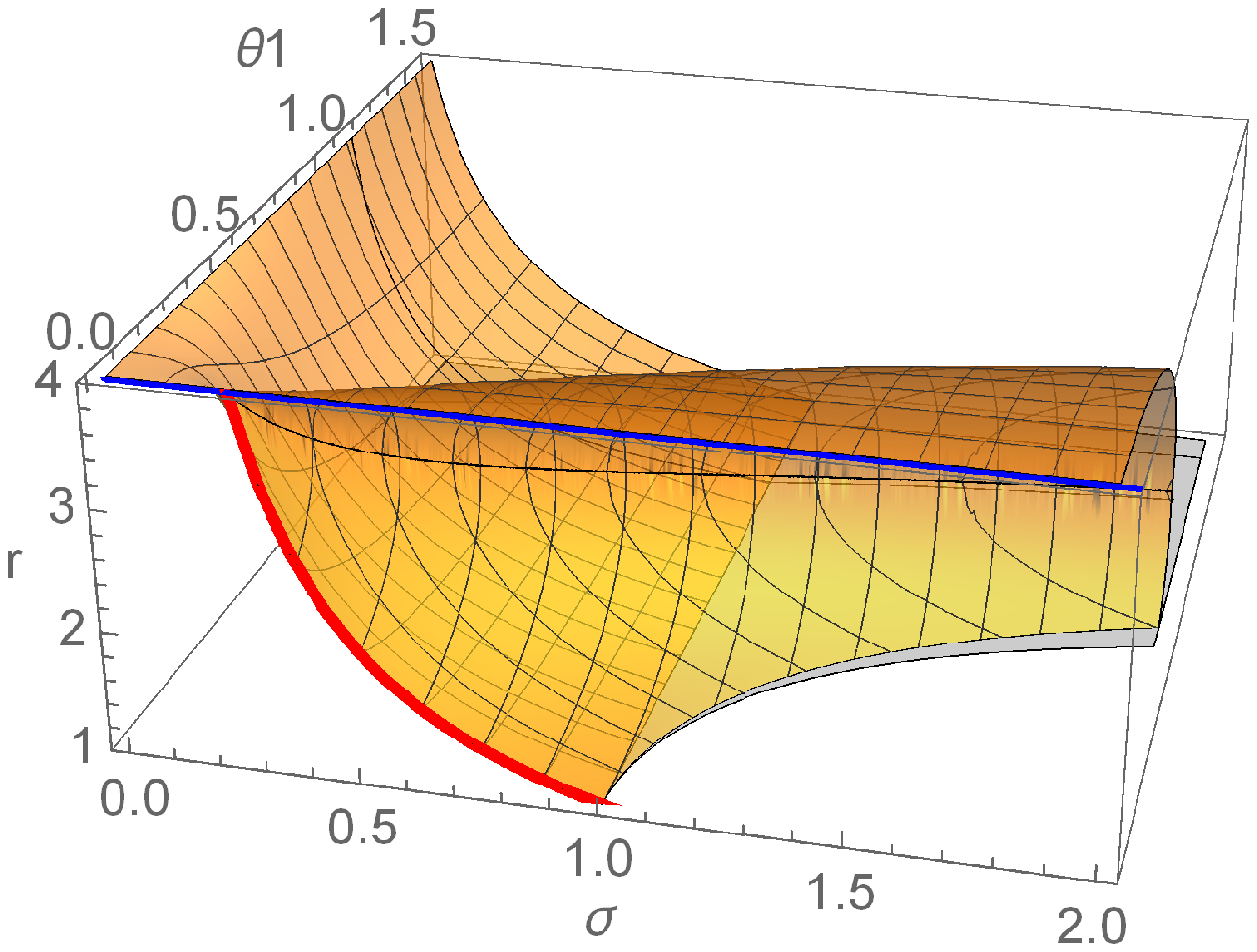}
  \caption{Left: Value of $\theta_2$ from Eq. (\ref{eq:theta2switchon}) in terms of $\sigma=\mathcal{M}_{A1}^{-2}$. Its maximum value is $\arccos\sqrt{2/5} \sim 0.56 \frac{\pi}{2}$.
 Right: MHD density jump $r$ in terms of $(\sigma,\theta_1)$ for $\theta_1 \in [0,\pi/2]$. Only the \textcolor[rgb]{0.00,0.07,1.00}{blue line}, which has $\theta_1=\theta_2=0$,  was considered in \cite{BretJPP2018}. The \rev{red line} is the switch-on solution (\ref{eq:rMHD}), with $\theta_1=0$ but $\theta_2\neq 0$.}\label{fig:theta2switchon}
\end{figure}

Equation (\ref{eq:1MHD}) clearly defines 2 kinds of shocks,
\begin{itemize}
  \item The first kind comes from $\tan \theta_2=0$, that is, $\theta_2=0$. Inserting it into (\ref{eq:2MHD}) gives $r=1$ or $r=4$. The first option, $r=1$ is the continuity solution, where nothing changes between the upstream and the downstream. The second option is the parallel shock solution, with $r=4$ for a sonic strong shock and an adiabatic index $\gamma=5/3$.
  \item Yet, (\ref{eq:1MHD}) also allows for,
  \begin{equation}\label{eq:rMHD}
  r = \mathcal{M}_{A1}^2,
  \end{equation}
  which is the MHD switch-on solution. Inserting it into (\ref{eq:2MHD}) gives,
  \begin{equation}\label{eq:theta2switchon}
  \cos^2\theta_2=\frac{3}{10 \mathcal{M}_{A1}^2-2 \mathcal{M}_{A1}^4-5}.
  \end{equation}
  The value of $\theta_2$ so defined is displayed on Fig. \ref{fig:theta2switchon}-left. $\theta_2 \neq 0$ is only permitted within a finite range of Alfv\'{e}n Mach numbers defined by $\cos^2\theta_2<1$, that is, $1 < \mathcal{M}_{A1} < 2$. Note that instead of parameterizing $\theta_2$ by the Alfv\'{e}n Mach number $\mathcal{M}_{A1}$, we choose the variable,
  \begin{equation}\label{eq:sigma}
  \sigma = \frac{B_1^2/4\pi}{m n_1 v_1^2} = \frac{1}{\mathcal{M}_{A1}^2}.
  \end{equation}
  This $\sigma$ parameter allows for a straightforward comparison with  PIC simulations where $\sigma$ is usually used instead of $\mathcal{M}_{A1}$ (see for example \cite{Sironi2011ApJ,BretApJ2020}). As a function of $\sigma$, $\theta_2 \neq 0$ is allowed for $\sigma \in [1/4,1]$.
\end{itemize}

For a finite upstream temperature $T_1 > 0$, MHD switch-on solutions are also restricted to a range of upstream temperatures via $\beta_1=n_1k_BT_1/B_1^2 < 2/\gamma$, where $\gamma$ is the adiabatic index \citep{Kennel1989,Sterck1999,Delmont2011}\footnote{See in particular Fig. 3 in \cite{Sterck1999}.}. Since the present work is limited to $T_1=0$, it cannot explore this dimension of the switch-on solutions range.

While they have been produced in the laboratory \citep{Craig1973}, such shocks have been rarely detected in space due to the smallness of the parameter window that allows them. \cite{Feng2009} reported the detection of a ``possible’’ interplanetary switch-on shock. Also, \cite{ Farris1994,Russell1995} reported the detection of one switch-on shock among the ISEE\footnote{International Sun-Earth Explorer, see \cite{ISEE1977}.} data. The more recent review of \cite{Balogh2013} still refers to \cite{Farris1994} as  ``the rare case of observation of a switch-on shock'' in its \S 2.3.6.

Finally, it is interesting to compute the MHD density jump $r$ for any upstream angle $\theta_1$. This can be done solving the ``shock adiabatic'' equation given in \cite{fitzpatrick} \S 7.21, and setting $T_1=0$\footnote{$V_{S1}=0$ in the notation of \cite{fitzpatrick}.}. The result is pictured on Fig. \ref{fig:theta2switchon}-right, in terms of $\theta_1 \in [0,\pi/2]$ and $\sigma$. Only the blue line, which has $\theta_1=\theta_2=0$,  was considered in \cite{BretJPP2018}. The red line is the switch-on solution (\ref{eq:rMHD}), with $\theta_1=0$ but $\theta_2\neq 0$.

\section{Method}\label{sec:Method}
Our method to determine the downstream anisotropy in terms of the upstream field relies on a monitoring of the kinetic history of the plasma as it crosses the front. In this process, the parallel and perpendicular temperatures of the plasma are changed according to some prescriptions explained below. The resulting state of the plasma downstream is labelled ``Stage 1''. Stage 1 is generally \emph{not} isotropic.

Depending on the strength of the downstream field $\mathbf{B}_2$, Stage 1 can be stable or not. If it is stable, then Stage 1 is the end state of the downstream. If it is unstable, then the plasma migrates towards its instability threshold, namely mirror or firehose stability. This is ``Stage 2''. In such case, Stage 2 is the end state of the downstream.

Stage 1 and 2 are therefore temporal evolving stages of the downstream plasma. This has been verified for the parallel case by the PIC simulations performed by \cite{Haggerty2022}, where the 2 stages have been clearly identified.

Also, the stability alluded here is not the one of the whole shock structure, like for example the corrugation instability (\cite{LandauFluid}, \S 90). It is rather the stability of the downstream plasma as an isolated and homogenous entity.

This algorithm was applied to the parallel and perpendicular cases in \cite{BretJPP2018} and \cite{BretPoP2019} respectively. In both cases, the orientation of $\mathbf{B}_2$ makes it simple to set the temperatures of Stage 1. We will now see that the obliquity of $\mathbf{B}_2$ demands further characterization of Stage 1.

\subsection{Characterization of Stage 1}
If the motion of the plasma through the front were adiabatic, the corresponding evolution of the parallel and perpendicular temperatures would be described by the double adiabatic equations  of \cite{CGL1956},
\begin{eqnarray}\label{eq:CGL}
   \frac{T_\parallel B^2}{n^2}&=&cst ,  \nonumber \\
     \frac{T_\perp}{B} &=& cst.
\end{eqnarray}
Here, like in the rest of the paper, the parallel and perpendicular are defined with respect to the local magnetic field.

Now, since we are dealing with shockwaves, the evolution of the plasma from the upstream to the downstream is \emph{not} adiabatic. For  parallel and  perpendicular shocks, this results in different prescriptions.
\begin{itemize}
  \item For the parallel shock case treated in \cite{BretJPP2018}, we took $\theta_{1,2}=\xi_{1,2}=0$ and considered the entropy excess goes into the parallel temperature. Intuitively, this steams from the fact that the transit of the plasma through the front can be viewed as a compression between 2 converging virtual walls, normal to the flow. These walls by no means exist. They are simply an analogy of how the entropy gain is realized.

      Regarding the perpendicular temperature, Eq. (\ref{eq:CGL}) simply gives $T_{\perp}=cst$ in the parallel case, since $B_2=B_1$ for such a shock. Such changes of the temperatures have been successfully checked through PIC simulations in \cite{Haggerty2022}.
  \item For the perpendicular shock case treated in \cite{BretPoP2019}, we took $\theta_{1,2}=\pi/2$ and  $\xi_{1,2}=0$. Here the plasma can still be viewed as compressed between 2 virtual walls normal to the flow. We therefore considered that the temperature normal to the flow, that is, parallel to the field, evolves adiabatically.
\end{itemize}

An additional constraint that must always be satisfied is the equality of the 2 temperatures perpendicular to the field, enforced by the Vlasov equation (\cite{LandauKinetic}, \S 53).

These considerations are summarized in Table \ref{tab:cases} which gives the values of $T_{2\parallel}$ and $T_{2\perp}$ in the parallel and perpendicular cases.

\begin{table}
\begin{center}
\begin{tabular}{l|c|c}
  Cases & $T_{2\parallel}$ & $T_{2\perp}$ \\
  \hline
  Parallel, $\theta_{1,2}=0$ & $T_1\left(\frac{n_2B_1}{n_1B_2}\right)^2$ + entropy & $T_1\frac{B_1}{B_2}$ \\
  Perpendicular, $\theta_{1,2}=\pi/2$ & $T_1\left(\frac{n_2B_1}{n_1B_2}\right)^2$ & $T_1\frac{B_1}{B_2}$ + entropy
\end{tabular}
\end{center}
 \caption{Values of $T_{2\parallel}$ and $T_{2\perp}$ in Stage 1 for the parallel and perpendicular cases.}\label{tab:cases}
\end{table}

As already stated in the introduction, MHD suggests that in a switch-on shock,  the obliquity $\theta_2$ of the downstream field can be as high as $0.56 \frac{\pi}{2}$. Hence, we need to interpolate between the 2 extremes of Table \ref{tab:cases}. We cannot just elaborate from \cite{BretJPP2018} by exploring $\theta_2 = \varepsilon$, with $0 < \varepsilon \ll 1$.

For intermediate values of $\theta_2$, we propose the following interpolation between the 2 extremes of Table \ref{tab:cases},
\begin{eqnarray}\label{eq:TS1}
  T_{2\parallel} =& T_1\left(\frac{n_2B_1}{n_1B_2}\right)^2    & + ~ T_e \cos^2\theta_2, \nonumber\\
  T_{2\perp}     =& T_1\frac{B_1}{B_2}                         & + ~ \frac{1}{2} T_e \sin^2\theta_2.
\end{eqnarray}

Our \emph{ansatz} is therefore that the downstream temperatures are the sum of the adiabatic ones given by \cite{CGL1956}, plus an entropy excess. For the parallel temperature, the entropy excess is a fraction $\cos^2\theta_2$ of a quantity we label $T_e$ (subscript ``e'' for \emph{e}ntropy). For the perpendicular temperature, the entropy excess is a fraction $\frac{1}{2}\sin^2\theta_2$ of the same $T_e$. Here, the factor $1/2$ accounts for the necessary identity of the 2 perpendicular temperatures. Finally, the 3 temperature excesses sum to $T_e$.

The $\cos^2\theta_2$ and $\sin^2\theta_2$ functions are the simplest choice fulfilling these requirements. Further works, notably PIC simulations (see conclusion), should allow to test their relevance.

Note that $T_e$ is not arbitrary but is solved for using the conservation equations (see Eq. (\ref{eq:Td}) in Appendix \ref{ap:1}). It represents the heat generated from the shock entropy.

We now compute the properties of Stage 1 accounting for these extended prescriptions for Stage 1.

\section{Properties of Stage 1}\label{sec:S1}
Due to the complexity of the calculations, we treat only the case of a sonic strong shock, namely $T_1=0$.

\subsection{Conservation equations for anisotropic temperatures}
The conservation equations for anisotropic temperatures in the downstream are established in Appendix \ref{ap:cons}. Though with different notations, they
 can be found in \cite{Hudson1970,Erkaev2000}. With $T_1=0$, they read,
\begin{eqnarray}
n_2 v_2 \cos \xi_2&=&n_1 v_1, \label{eq:1}  \\
B_2 \cos \theta_2&=&B_1 ,       \label{eq:2}  \\
B_2 v_2 \sin \theta_2 \cos \xi_2 - B_2 v_2 \cos \theta_2 \sin \xi_2&=& 0,      \label{eq:3}  \\
\cos ^2\theta_2 n_2 k_B T_{2 \parallel}+ \sin ^2\theta_2 n_2 k_B T_{2 \perp}+ m n_2 v_2^2 \cos ^2\xi_2 -\frac{B_2^2 \cos \left(2 \theta _2\right)}{8 \pi } &=&  -\frac{B_1^2 }{8 \pi } +m n_1 v_1^2  , \label{eq:4} \\
\sin \theta_2 \cos \theta_2 n_2 k_B\left(T_{2 \parallel}-T_{2 \perp}\right)+m n_2 v_2^2 \sin \xi_2 \cos \xi_2-\frac{B_2^2 \sin \left(2 \theta _2\right)}{8 \pi }&=& 0 ,  \label{eq:5}  \\
\left[v \left(\mathcal{A} \cos \xi+\mathcal{B} \cos \xi+\mathcal{C} \sin \xi\right)\right]_1^2&=& 0, \label{eq:6}
\end{eqnarray}
where,
\begin{eqnarray}
\mathcal{A}&=&\frac{1}{2} n k_B T_{\parallel}+n k_B T_{\perp}+\frac{B^2}{8 \pi }+\frac{1}{2} m n v^2, \nonumber  \\
\mathcal{B}&=&-\frac{B^2 \cos \left(2 \theta \right)}{8 \pi }+ \cos^2\theta ~ n k_B T_{\parallel}+ \sin^2\theta ~ n k_B T_{\perp}, \nonumber  \\
\mathcal{C}&=& \sin \theta \cos \theta ~ n k_B\left(T_{\parallel}-T_{\perp}\right)-\frac{B^2 \sin \left(2 \theta \right)}{8 \pi }. \nonumber
\end{eqnarray}
In equation (\ref{eq:6}), the notation $\left[Q\right]_1^2$ stands for the difference of any quantity $Q$ between the upstream and the downstream.

With $T_1=0$, prescriptions (\ref{eq:TS1}) for the downstream temperatures in Stage 1 simply read,
\begin{eqnarray}\label{eq:TS1T10}
  T_{2\parallel} &=&  T_e \cos^2\theta_2, \\
  T_{2\perp} &=& \frac{1}{2} T_e \sin^2\theta_2. \nonumber
\end{eqnarray}

\subsection{Resolution of the system of equations}\label{sec:S1reso}
The resolution of the system (\ref{eq:1}-\ref{eq:TS1T10}) is lengthy and reported in Appendix \ref{ap:1}. It turns out that it is convenient to determine first the angle $\theta_2$ as a function of $\sigma$, and then to compute the density jump $r(\sigma)$.

\begin{figure}
\begin{center}
 \includegraphics[width=0.5\textwidth]{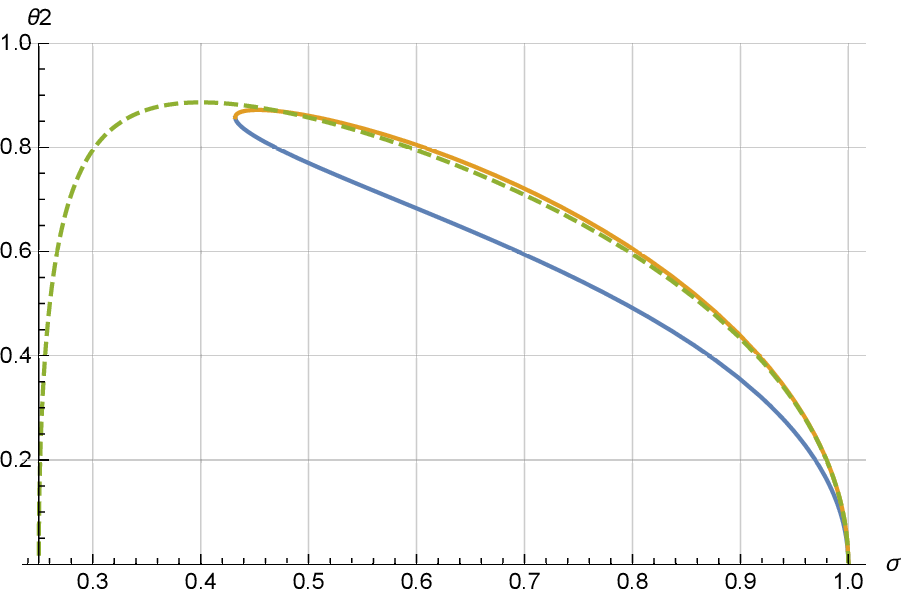}\includegraphics[width=0.5\textwidth]{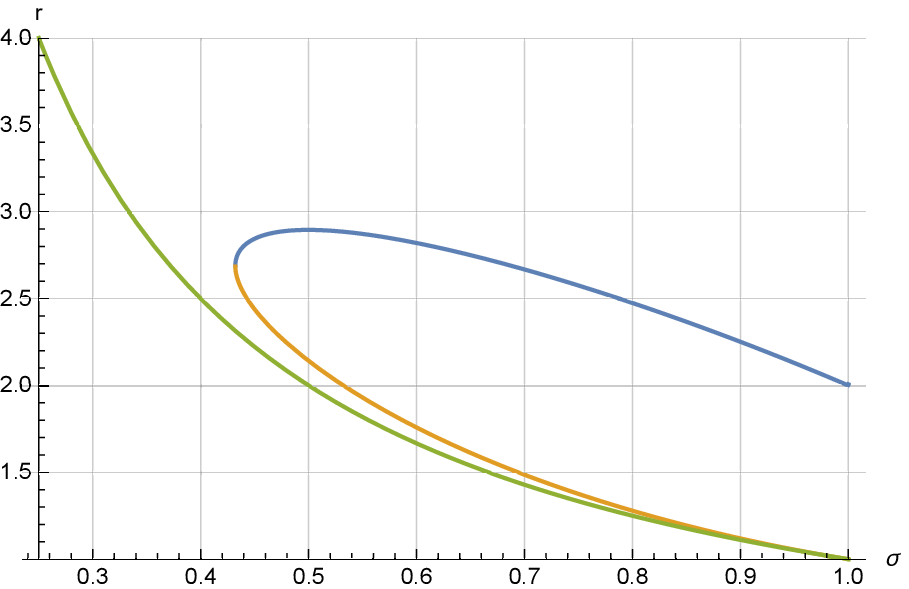}
\end{center}
\caption{Left: Values of $\theta_2$ arising from $Q=0$ in Eq. (\ref{eq:Q}). Solutions exist only for $\sigma \in [0.432, 1]$. Right: Corresponding values of the density jump from Eq. (\ref{eq:rS1}). The color code refers to the corresponding $\theta_2$-branch. The \textcolor[rgb]{0.00,0.78,0.00}{green curves} pertain to the MHD switch-on solution.}\label{fig:theta2}
\end{figure}

The algebra unravels 3 $\theta_2$-branches for $\theta_1=0$,
\begin{itemize}
  \item One branch is simply $\theta_2=0$, with $r=1$ and $r=2$. The first one, with $r=1$, is the continuity solution. The second one, with $r=2$, is the parallel strong sonic shock solution for Stage 1, already studied in \cite{BretJPP2018}.
  \item The other branch defines 2 values of $\theta_2(\sigma)$ which correspond to our switch-on solutions. They are pictured on Figure \ref{fig:theta2}-left. Then the corresponding density jump $r(\sigma)$ is computed and plotted on Figure \ref{fig:theta2}-right. The green curve pictures the MHD switch-on solution Eq. (\ref{eq:rMHD}), defined for $\sigma \in [1/4,1]$. In Stage 1, numerical exploration shows solutions exists only for $\sigma \in [0.432, 1]$.

     Therefore, while there is only 1 switch-on solution in MHD, our model offers 2. Figure \ref{fig:theta2}-left shows that both of our branches merge with the MHD result for $\sigma=1$ as far as $\theta_2$ is concerned. Such is only the case for the lower of our $r$-branch, as can be seen from Figure \ref{fig:theta2}-right.
\end{itemize}

Figure \ref{fig:theta2}-left shows that the largest value of $\theta_2$ in Stage 1 is almost as high as its MHD counterpart.

In accordance with the method explained in Section \ref{sec:Method}, we now  study the stability of Stage 1.

\subsection{Stability of Stage 1}\label{sec:Stab}
If unstable, Stage 1 is mirror or firehose unstable. The thresholds for these instabilities are given by \citep{Gary1993,Gary2009},
\begin{equation}\label{eq:tresholds}
 \frac{T_{2\perp}}{T_{2\parallel}} \equiv A_2 = 1 \pm \frac{1}{\beta_{\parallel 2}},
\end{equation}
where
\begin{equation}\label{eq:beta2para}
\beta_{\parallel 2} = \frac{n_2  k_B T_{2\parallel}}{B_2^2/(8\pi)},
\end{equation}
and the ``+'' and ``-'' signs stand for the thresholds of the mirror and firehose instabilities respectively. From Eq. (\ref{eq:TS1T10}), we obtain the downstream anisotropy $A_2$,
\begin{equation}\label{eq:A2}
A_2 = \frac{1}{2} \tan^2\theta_2.
\end{equation}
For $\beta_{\parallel 2}$, we obtain in Appendix \ref{ap:beta},
\begin{equation}\label{eq:beta2}
\beta_{\parallel 2} = - 2 \frac{r \sigma \tan^2\theta_2-2 r+2}{  r \sigma  (\tan ^4\theta_2+2 ) }.
\end{equation}

\begin{figure}
\begin{center}
 \includegraphics[width=0.5\textwidth]{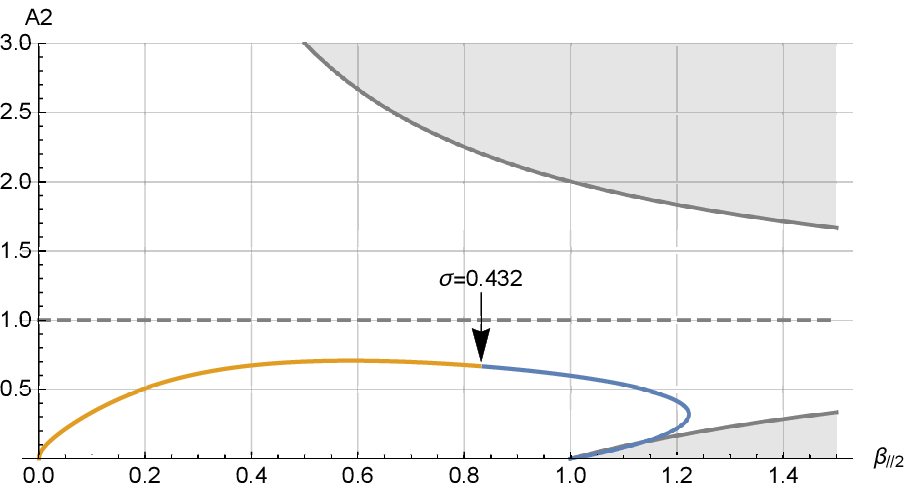}\includegraphics[width=0.5\textwidth]{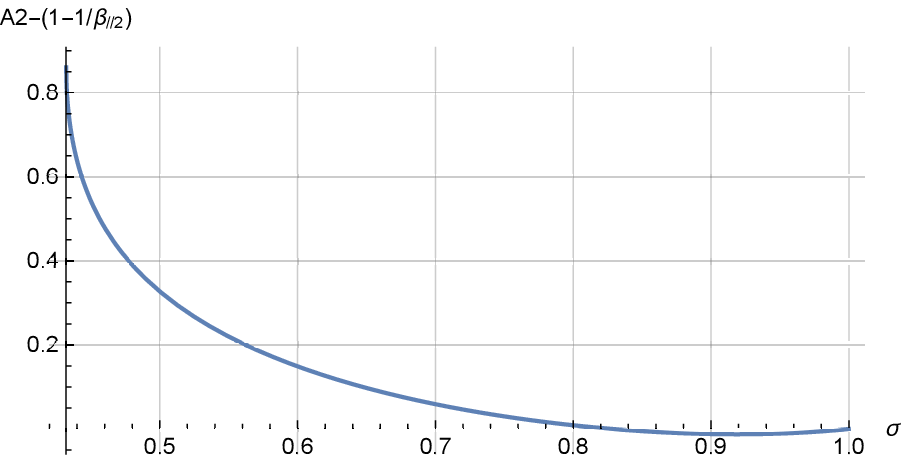}
\end{center}
\caption{Left: Thresholds for the mirror (upper gray line) and firehose (lower gray line) instabilities. The plasma is unstable within the shaded areas. The loop shows the curves $(\beta_{\parallel 2}(\sigma),A_2(\sigma))$ for the 2 $\theta_2$-branches defined previously. The 2 branches start from the same point for $\sigma=0.432$, and both reach $A_2=0$ for $\sigma=1$. Right: Stability analysis of the blue branch for $\sigma \in [0.432, 1]$. It is found firehose unstable for $\sigma \in [0.82, 1]$.}\label{fig:StabS1}
\end{figure}

In order to assess the stability of Stage 1, we then proceed as follow,
\begin{itemize}
  \item From Eq. (\ref{eq:tresholds}), we plot the thresholds for the mirror and firehose instabilities in the $(\beta_{\parallel 2},A_2)$ plane.
  \item Then on the same graph, we plot the curves $(\beta_{\parallel 2},A_2)$ for the 2 non-trivial $\theta_2$-branches found in Section \ref{sec:S1reso}.
\end{itemize}
The result is pictured on Figure \ref{fig:StabS1}-left. In \cite{BretJPP2018}, Stage 1 had $A_2=0$ for the sonic strong shock case. Here, $A_2$ departs from 0 but remains small.

It turns out that the orange branch pictured on Figure \ref{fig:theta2}-left, namely the one closest to the MHD solution, is stable for any  $\sigma$. Yet, the blue one is slightly unstable in some $\sigma$ range. For this branch, the quantity $A_2 - (1-\beta_{\parallel 2}^{-1})$ is plotted on Figure \ref{fig:StabS1}-right. It is negative for $\sigma \in [0.83, 1]$, indicating firehose instability. In this $\sigma$-range, the downstream will therefore migrate to Stage 2, on the firehose threshold.

As can be seen on Figure \ref{fig:StabS1}-left, the blue branch Stage 1 is only slightly unstable. Consequently, the corresponding marginally stable Stage 2 is very close to the unstable states. This is confirmed below in Section \ref{sec:S2}, where Stage 2 is analysed.

For now, in order to document the differences between our 2 branches, we further study Stage 1 by computing its entropy and its Alfv\'{e}nic downstream velocity.

\subsection{Entropy of Stage 1}\label{sec:S}
The 2 branches for Stage 1 cannot be distinguished by their energy since they both fulfill the energy conservation equation (\ref{eq:6}), where the upstream energy is the same in both cases. Their energy densities are therefore identical. Yet, they can be distinguished on the basis of their entropy.

For a bi-Maxwellian of the form,
\begin{equation}\label{eq:Max}
F= \frac{n}{\pi^{3/2} \sqrt{a} b}\exp \left(-\frac{v_x^2}{a}\right) \exp \left(-\frac{v_y^2+v_z^2}{b}\right),
\end{equation}
where $a=2k_BT_\parallel/m$ and $b=2k_BT_\perp/m$,
the entropy reads,
\begin{equation}\label{eq:entropy}
S = -k_B \int F \ln F d^3v = \frac{1}{2} k_B n \left[ 3 + \ln (\pi^3 a b^2 )-2 \ln n\right],
\end{equation}
where $n=\int F d^3v$. Using the subscript ``b’’ for the blue branch on Fig. \ref{fig:theta2}, and subscript ``o’’ for the orange one, we get for the entropy difference per particle between the 2 branches,
\begin{eqnarray}\label{eq:entropydiff}
\frac{2}{k_B} \left(\frac{S_o}{n_o}-\frac{S_b}{n_b}\right) \equiv \Delta s &=&  \ln \left[  \frac{a_o b_o^2}{a_b b_b^2} \right]
                        + 2\ln \frac{n_b}{n_o},  \nonumber  \\
&=&   \ln \left[  \frac{T_{\parallel 2,o} ~ T_{\perp 2,o}^2}{T_{\parallel 2,b} ~ T_{\perp 2,b}^2} \right]+2\ln \frac{n_b}{n_o} ,  \nonumber  \\
&=&   \ln \left[  \frac{T_{\parallel 2,o}^3 ~ A_{2,o}^2}{T_{\parallel 2,b}^3 ~ A_{2,b}^2} \right]+2\ln \frac{n_b}{n_o}, \nonumber  \\
&=&   \ln \left[ \left( \frac{T_{e,o} \cos^2\theta_{2,o}}{ T_{e,b} \cos^2\theta_{2,b} }  \right)^3 \frac{A_{2,o}^2}{A_{2,b}^2} \right] + 2\ln \frac{n_b}{n_o},
\end{eqnarray}
where we have used $A_2=T_{\perp 2}/T_{\parallel 2}$ and then $T_{\parallel 2} = T_e \cos^2\theta_2$ for both branches.

\begin{figure}
\begin{center}
 \includegraphics[width=0.5\textwidth]{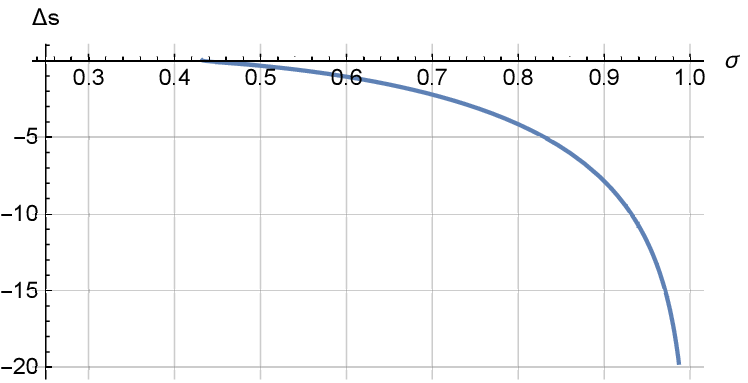}\includegraphics[width=0.5\textwidth]{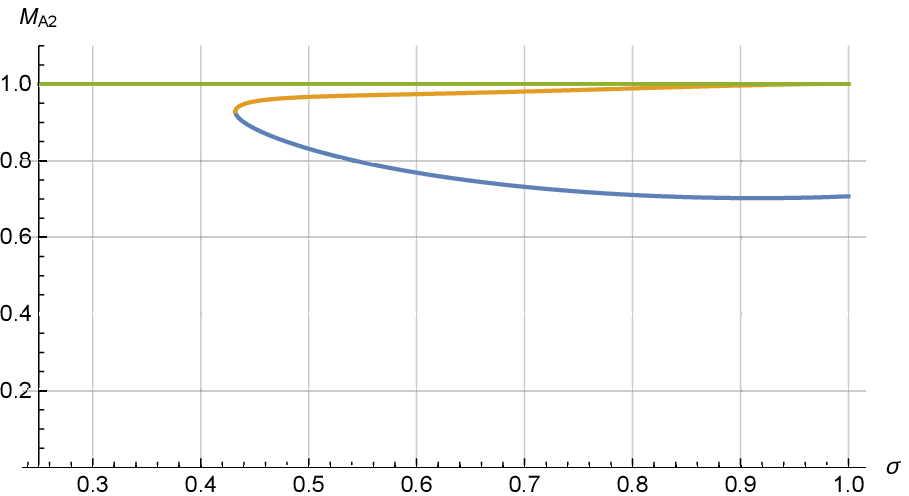}
\end{center}
\caption{Left: Entropy difference $\Delta s$ as defined by Eq. (\ref{eq:entropydiff}), between the blue branch on Fig. \ref{fig:theta2}, and the orange one. $\Delta s < 0$ implies the orange branch has lower entropy than the blue one. Right: Value of $\mathcal{M}_{A2}$ in MHD (green) and for the 2 branches of our model.}\label{fig:entropy}
\end{figure}

The numerical evaluation\footnote{$T_e$ is given by Eq. (\ref{eq:Td}).} of this quantity displayed on Figure \ref{fig:entropy}-left shows that $\Delta s < 0$ for any $\sigma \in [0.432, 1]$. Therefore, $S_o/n_o < S_b/n_b$ for any $\sigma$. The orange branch on Fig. \ref{fig:theta2} has \emph{lower} entropy than the blue one.

\subsection{Downstream Alfv\'{e}nic Mach number of Stage 1}\label{sec:MA2}
Another difference between the 2 branches lies in their respective Alfv\'{e}nic Mach number, namely,
\begin{equation}\label{eq:MA2}
\mathcal{M}_{A2}^2 = \frac{m n_2 v_2^2}{B_2^2/4\pi}.
\end{equation}
From Eq. (\ref{eq:2}) we get $B_2=B_1/\cos\theta_2$.  Then Eq. (\ref{eq:1}) gives $v_2 = n_1v_1/n_2\cos\xi_2 = n_1v_1/n_2\cos\theta_2$, since $\xi_2=\theta_2$ in our model as in MHD\footnote{See Eq. (\ref{eq:mhd3_1}) for MHD and Appendix \ref{ap:1} for our model.}. We finally obtain, in terms of the dimensionless variables (\ref{eq:dimless}), in both our model and MHD,
\begin{equation}\label{eq:MA2OK}
\mathcal{M}_{A2} = \frac{\mathcal{M}_{A1}}{\sqrt{r}} = \frac{1}{\sqrt{\sigma r}}.
\end{equation}
\begin{itemize}
  \item In MHD, Eq. (\ref{eq:rMHD})  has $r=\mathcal{M}_{A1}^2$, so that for the switch-on MHD shock, $\mathcal{M}_{A2} = 1$ (\cite{Goedbloed2010}, p. 853).
  \item In our model, the value of $\mathcal{M}_{A2}$ is pictured on Fig. \ref{fig:entropy}-right for the 2 branches represented on Fig. \ref{fig:theta2}. Our 2 branches are found slightly sub-Alfv\'{e}nic.
\end{itemize}

\section{Properties of Stage 2}\label{sec:S2}
The firehose instability of the blue branch for $\sigma \in [0.83,1]$ requires studying the properties of Stage 2 when marginally firehose stable. The conservation equations are the same. But instead of imposing prescriptions (\ref{eq:TS1}) for the temperatures, we now impose firehose marginal stability for the downstream, namely,
\begin{equation}\label{eq:tresholdsfire}
 \frac{T_{2\perp}}{T_{2\parallel}} = 1 - \frac{1}{\beta_{\parallel 2}}.
\end{equation}

The resolution of the system follows the same path as that describes in Appendix \ref{ap:1} for Stage 1. It yields 3 equations for $\theta_2,\xi_2$ and $r$,
\begin{eqnarray}
  \tan \theta_2-\tan \xi_2 &=& 0, \label{eq:S21}\\
  \frac{2 \tan \xi_2}{r}-\frac{\tan \theta_2}{\mathcal{M}_{A1}^2} &=& 0, \label{eq:S22}\\
  \mathcal{M}_{A1}^2 \left(-\sec ^2\xi_2+(r-5) r+5\right)+r \tan \theta_2 \tan \xi_2+r &=& 0. \label{eq:S23}
\end{eqnarray}
Equation (\ref{eq:S21}) imposes again $\theta_2 = \xi_2$. Replacing in Eq. (\ref{eq:S22}) gives,
\begin{equation}
\tan \theta_2 ( r - 2\mathcal{M}_{A1}^2  ) = 0,
\end{equation}
which leaves 2 options only,
\begin{itemize}
  \item $\theta_2=0$, which pertains to the parallel shock solution. Setting then $\theta_2=0$ in Eq. (\ref{eq:S23}) then gives exactly the Stage 2 solution found in \cite{BretJPP2018}\footnote{See Eq. (3.5) of \cite{BretJPP2018} for $\chi_1=\infty$.}.
  \item The other option is,
  \begin{equation}\label{eq:rS2}
  r=2\mathcal{M}_{A1}^2=2/\sigma.
  \end{equation}
  Inserting this result in Eq. (\ref{eq:S23}) and solving for $\theta_2$ gives,
  \begin{equation}
     \cos^2 \theta _2 = \frac{1}{10 \mathcal{M}_{A1}^2 - 4 \mathcal{M}_{A1}^4   - 5},
  \end{equation}
  reminiscent of Eq. (\ref{eq:theta2switchon}) for the MHD case. Solutions can here be found for,
  \begin{eqnarray}
     \mathcal{M}_{A1} & \in & \left[ \frac{\sqrt{5-\sqrt{5}}}{2} , \frac{\sqrt{5+\sqrt{5}}}{2} \right] \sim [0.83, 1.34] , \nonumber \\
     \Leftrightarrow  \sigma = \frac{1}{\mathcal{M}_{A1}^2}       & \in  &  \left[ \frac{4}{5+\sqrt{5}}, \frac{4}{5-\sqrt{5}} \right] \sim  [0.55, 1.44]. \label{eq:S2range}
  \end{eqnarray}

\end{itemize}
The counterpart to switch-on shock is therefore recovered in our model for Stage 2 as well, still in a limited range of Alfv\'{e}n Mach numbers.

\begin{figure}
\begin{center}
 \includegraphics[width=\textwidth]{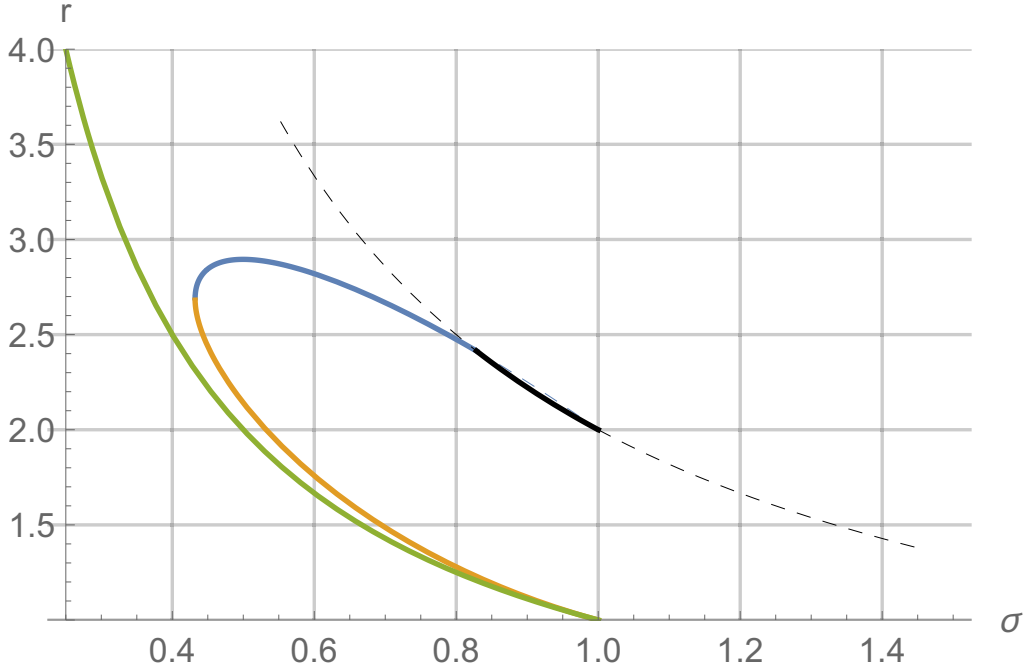}
\end{center}
\caption{Same as Figure \ref{fig:theta2}-right but showing how Stages 1 \& 2 fit together when accounting for the firehose instability of the blue branch for $\sigma \in [0.83, 1]$. Stage 2 density jump $r=2/\sigma$ is discontinued from $\sigma = 1$ since Stage 1 has no solution in this range.}\label{fig:rfinal}
\end{figure}

Figure \ref{fig:rfinal} is eventually the end result of the present work. Like Figure \ref{fig:theta2}-right, it features the density jump of the MHD switch-on solution, together with the 2 branches of our model. But here, the way Stages 1 \& 2 fit together in the $\sigma$ unstable range is elucidated. Since the blue branch has been found firehose unstable for $\sigma \in [0.83, 1]$, it is replaced by Stage 2, namely Eq. (\ref{eq:rS2}), in this range. As expected, the corresponding density jump is very close to the one of Stage 1 since the system is almost marginally stable in this range, while Stage 2 sits exactly on marginal stability.

On Figure \ref{fig:rfinal}, the jump for Stage 2, namely $r=2/\sigma$, is showed in black and plotted within the full range (\ref{eq:S2range}) where it is defined. For $\sigma < 0.83$, the line is dashed because Stage 1 is stable, hence defining the density jump. Then for $\sigma \in [0.83, 1]$, the blue branch is dashed since it pertains to the unstable Stage 1. There, the jump is now given by Stage 2 through $r=2/\sigma$. Beyond $\sigma = 1$, Stage 1 offers no solutions. Since in our scenario Stage 1 is the first state of the downstream after crossing the front, the shock cannot accommodate such values of $\sigma$ in the switch-on regime. For $\sigma > 1$, there is therefore no Stage 1 from where the system could jump to Stage 2, even though Stage 2 offers solutions. As a consequence, the black curve is dashed from $\sigma =1$ to $1.44$.

\section{Conclusion}\label{sec:conclu}
In a collisionless non-magnetized plasma, the Weibel instability ensures isotropy, since it makes anisotropies unstable \citep{Weibel,SilvaPRE2021}. Therefore, for collisionless shocks in such medium, the only source of departures from MHD should stem from accelerated particles \citep{HaggertyApJ2020,BretApJ2020}.

In contrast, a temperature anisotropy can be stabilized in a collisionless plasma by an external magnetic field. Therefore, if the plasma turns anisotropic when crossing the front of a collisionless shock, its downstream anisotropy could be stable, resulting in a departure from MHD.

Several authors studied the conservation equations for a shock accounting for anisotropic pressures. Yet, the downstream degree of anisotropy is considered a free parameter in these works \citep{Erkaev2000,Double2004,Gerbig2011}. In the present article, we devised a model allowing to compute the degree of anisotropy of the downstream, in terms of the upstream parameters. We focused on the switch-on solutions where the field is aligned with the flow in the upstream, but not in the downstream.

For such a configuration, MHD allows for one shock solution, the switch-on solution, for which the density jump is given by Eq. (\ref{eq:rMHD}). According to our model, which has been successfully tested against PIC simulations for the parallel case \citep{Haggerty2022}, there are two collisionless switch-on solutions for which the angle $\theta_2$ and the density jump $r$ are plotted  on Figures \ref{fig:theta2} \& \ref{fig:rfinal}.One solution for what we named ``Stage 1'' is stable for any $\sigma$ where it is defined. The other is slightly firehose unstable within a limited $\sigma$-range. Exploring then Stage 2 in this range allows to correct the computed density jump. Since the Stage 1 that needed to be corrected was only slightly firehose unstable, the correction found with Stage 2 \emph{marginally} firehose stable, is small.

The existence of 2 switch-on solutions in our model instead of 1 in MHD could be explained. We plotted on Figure \ref{fig:theta2switchon}-right the MHD solutions for a cold upstream and any upstream field obliquity $\theta_1$. One can see that the MHD switch-on solution for $\theta_1=0$ splits into 2 different solutions for $\theta_1 > 0$. These 2 solutions are the intermediate and fast shocks. They merge for $\theta_1=0$, which is why MHD switch-on shocks can be termed intermediate or fast (\cite{Goedbloed2010}, p. 853).

Possibly within our model, these 2 kinds of shocks do not merge for $\theta_1=0$. Future works dealing with the fully oblique case $\theta_1 > 0$ will explore how the MHD intermediate and fast shocks morph within our model.

Is one of our 2 branches physically favored? Both pertain to a downstream plasma with the same energy density since both fulfil the energy conservation equation (\ref{eq:6}) where the upstream term is the same. We see from Figs. \ref{fig:theta2} \& \ref{fig:entropy} that the orange one is the closest to the MHD solution, yet we found in Section \ref{sec:S} that it has lower entropy than the blue branch. Further works, notably PIC simulations, would be needed to find out if these 2 branches are just our model's version of the oblique intermediate and fast shocks in the limit $\theta_1=0$.

In the same way that the theory devised for the parallel case has been tested through PIC simulation \citep{BretJPP2018,Haggerty2022}, it would be interesting to test the present conclusions through the same means. Yet, to our knowledge, no PIC simulations of switch-on shocks have been performed to date \citep{sironi_private}. An option in this respect would be to reproduce in PIC the bow shock MHD simulation performed in \cite{Sterck1999}. There, it was found that a portion of the bow shock produced by the simulation was of the switch-on type. Possibly a PIC counterpart of this work would allow to produce a switch-on shock and study it at the kinetic scale.

\section*{Acknowledgments}
Thanks are due to Lorenzo Sironi and Bertrand Lemb\`{e}ge for enriching discussions.

\section*{Funding}
A.B. acknowledges support by grants ENE2016-75703-R from the Spanish Ministerio
de Econom\'{\i}a y Competitividad and SBPLY/17/180501/000264 from the Junta de
Comunidades de Castilla-La Mancha. R.N. acknowledges support from the NSF Grant No. AST-
1816420. R.N. thanks the Black Hole Initiative at Harvard University for support. The BHI is funded by grants from the John Templeton Foundation and the Gordon and Betty Moore Foundation.

\section*{Declaration of Interests}
The authors report no conflict of interest.

\appendix

\section{Derivation of the conservation equations for anisotropic temperatures}\label{ap:cons}
Equations (\ref{eq:1}-\ref{eq:3}) are identical to their MHD counterpart since they do not involve the pressure. The differences due to anisotropic pressure are rather to be found in Eqs. (\ref{eq:4}-\ref{eq:6}). For Eqs. (\ref{eq:4},\ref{eq:5}), we start from the momentum flux density tensor equation (\cite{LandauFluid}, \S 7),
\begin{equation}
\frac{\partial (\rho v_i)}{\partial t} = - \frac{\partial \Pi_{ik}}{\partial x_k}.
\end{equation}
In the shock frame, the left-hand-side vanishes. Using the basis $x,y,z$ represented on Fig. \ref{fig:system}, where the shock jump is in the $x$ direction, we obtain the following jump conditions,
\begin{equation}\label{eq:Pis}
\left[ \Pi_{xx} \right]_1^2 = \left[ \Pi_{xy} \right]_1^2 = 0,
\end{equation}
where the notation $\left[Q\right]_1^2$ stands for the difference of any quantity $Q$ between the upstream and the downstream.

There are 3 contributions to the tensor $\Pi_{ik}$: ram pressure, magnetic pressure and thermal pressure:
\begin{itemize}
  \item The ram pressure part reads,
  \begin{equation}\label{eq:ram}
    \Pi_{ram} = \left( \begin{array}{ccc}
                         nmv^2\cos^2\xi       & nmv^2\cos\xi\sin\xi & 0 \\
                         nmv^2\cos\xi\sin\xi  & nmv^2\sin^2\xi      & 0 \\
                         0 & 0 & 0
                       \end{array}
     \right),
  \end{equation}
  where all quantities are to be taken with subscript 1 for the upstream and 2 for the downstream.
  \item For the magnetic pressure, we start in a basis $(x',y',z')$ aligned with the field. In such a basis,
    \begin{equation}
    \Pi_{mag}' = \left( \begin{array}{ccc}
                         -B^2/8\pi       & 0 & 0 \\
                         0  & B^2/8\pi   & 0 \\
                         0 & 0 & B^2/8\pi
                       \end{array}
     \right).
  \end{equation}
  We now express this tensor in our basis $(x,y,z)$, where $z=z'$ and $(x',y')$ are rotated by an angle $\theta$. Hence, we compute $R^{-1}\Pi_{mag}'R$, with
\begin{equation}\label{eq:rot}
R=\left(
\begin{array}{ccc}
 \cos \theta  & \sin \theta  & 0 \\
 -\sin \theta  & \cos \theta & 0 \\
 0 & 0 & 1 \\
\end{array}
\right).
\end{equation}
The result is,
  \begin{equation}\label{eq:mag}
    \Pi_{mag} =  \frac{B^2}{8\pi}\left( \begin{array}{ccc}
                         -\cos 2\theta   & -\sin 2\theta  & 0 \\
                         -\sin 2\theta   & \cos 2\theta       & 0 \\
                         0 & 0 & 1
                       \end{array}
     \right).
  \end{equation}
  \item The calculation is similar for the thermal pressure. We start in a basis adapted to the field,
  \begin{equation}
\Pi_{th}' = nk_B\left(
\begin{array}{ccc}
 T_{\parallel} & 0 & 0 \\
 0 & T_{\perp} & 0 \\
 0 & 0 & T_{\perp} \\
\end{array}
\right),
\end{equation}
  where the directions $\parallel$ and $\perp$ are considered with respect to the field. Computing $R^{-1}\Pi_{th}'R$, where the tensor $R$ is still given by Eq. (\ref{eq:rot}), gives in our basis $(x,y,z)$,
    \begin{equation}\label{eq:th}
  \Pi_{th} = nk_B\left(
\begin{array}{ccc}
 T_{\parallel}\cos^2\theta + T_{\perp}\sin^2\theta & (T_{\parallel}-T_{\perp}) \cos\theta\sin\theta & 0 \\
 (T_{\parallel}-T_{\perp}) \cos\theta\sin\theta    & T_{\parallel}\sin^2\theta + T_{\perp}\cos^2\theta & 0 \\
 0 & 0 & T_{\perp} \\
\end{array}
\right).
\end{equation}
\end{itemize}
When adding the contributions (\ref{eq:ram},\ref{eq:mag},\ref{eq:th}), the conservation equations (\ref{eq:Pis}) yield Eqs. (\ref{eq:4},\ref{eq:5}).

For the last equation, namely Eq. (\ref{eq:6}), we start from the energy conservation equation (\cite{LandauFluid}, \S 6),
\begin{equation}\label{eq:ener}
\frac{\partial}{\partial t}\left(\frac{1}{2}nmv^2+\varepsilon_{mag} + \varepsilon_{th} \right)
 = -\frac{\partial}{\partial x_k}\left[ v_k \left(\frac{1}{2}nmv^2+\varepsilon_{mag} + \varepsilon_{th} \right) \right]
 -\frac{\partial}{\partial x_k}\left[ v_i \left( \Pi_{ik,mag} + \Pi_{ik,th}  \right) \right],
\end{equation}
where $\varepsilon$ is the internal energy density,
\begin{eqnarray}
  \varepsilon_{mag} &=& \frac{B^2}{8\pi}, \nonumber\\
  \varepsilon_{th} &=&  \frac{1}{2} n k_BT_\parallel + n k_BT_\perp, \nonumber
\end{eqnarray}
and $\Pi_{mag} , \Pi_{th}$ are given by Eqs. (\ref{eq:mag},\ref{eq:th}) respectively. Setting the left-hand-side of Eq. (\ref{eq:ener}) to 0, and equating the right-hand-side between the upstream and the downstream, gives Eq. (\ref{eq:6}).

\section{Resolution of the system (\ref{eq:1}-\ref{eq:TS1T10})}\label{ap:1}

The system (\ref{eq:1}-\ref{eq:TS1T10}) can be reduced to a system of 3 equations for $r\equiv n_2/n_1, \theta_2$ and $T_e$. The pathway to do so is,
\begin{itemize}
  \item We first notice that equation (\ref{eq:3}) imposes $\xi_2=\theta_2$ (as in MHD). We can therefore set $\xi_2=\theta_2$ everywhere.
  \item We then use equation (\ref{eq:1}) to eliminate $v_2$ everywhere.
  \item Next, we use equation (\ref{eq:2}) to eliminate $B_2$ everywhere.
\end{itemize}

At this junction we are left with 3 equations which are the updated versions of (\ref{eq:4}-\ref{eq:6}). They read,
\begin{eqnarray}
r^2 \overline{T_e} \sin ^4\theta_2+2 r^2 \overline{T_e} \cos ^4\theta_2+r \sigma \tan ^2\theta_2-2 r+2  &=& 0, \label{eq:41}\\
r^2 \overline{T_e} \left[2 \sin (2 \theta_2)+3 \sin (4 \theta_2)\right]+16 \tan \theta_2 (1-r \sigma)  &=& 0, \label{eq:51}\\
\sec ^2\theta_2+r^2 \left[\overline{T_e} \cos (2 \theta_2)+2 \overline{T_e}-1 \right]  &=&  0, \label{eq:61}
\end{eqnarray}
in terms of the density ratio $r$ and the magnetic parameter $\sigma$ defined in Eqs. (\ref{eq:sigma}), plus,
\begin{equation}
\overline{T_e} = \frac{k_B T_e}{m v_1^2}.
\end{equation}

For further progress, it is convenient to define,
\begin{equation}
X_2 = \arcsin \theta_2.
\end{equation}
This change of variables makes the forthcoming equations polynomial in $X_2$, easy to solve numerically. The value of $\overline{T_e}$ can be extracted from equation (\ref{eq:41}) and reads,
\begin{equation}\label{eq:Td}
\overline{T_e} = \frac{r [(\sigma+2) X_2^2-2]-2 X_2^2+2}{r^2 (X_2^2-1) (X_2^2 (3 X_2^2-4)+2)}.
\end{equation}
Substituting it in Eqs. (\ref{eq:51},\ref{eq:61}) yields the 2 equations,
\begin{eqnarray}
X_2 \underbrace{\left[r \left\{\sigma (4 + 3 (X_2^2-2) X_2^2)-6 X_2^4+10 X_2^2-4\right\}-2 X_2^2\right]}_{\equiv \Lambda} &=& 0, \label{eq:52}\\
\sum_{k=0}^3 a_k X_2^{2k}, &=&  0, \label{eq:62}
\end{eqnarray}
with,
\begin{eqnarray}
  a_0 &=& 2 (r-2) (r-1), \nonumber\\
  a_1 &=& r (-6 r+3 \sigma+10)-6, \nonumber\\
  a_2 &=& 7 r^2-2 (\sigma+2) r+1 ,\nonumber\\
  a_3 &=&  -3 r^2.
\end{eqnarray}
Equation (\ref{eq:52}) clearly displays 2 branches,
\begin{itemize}
  \item One branch is $X_2 = 0$, that is, $\theta_2=0$. Inserting it into (\ref{eq:62}) gives $a_0=0$, that is, $r=1$ or $r=2$. The first one, with $r=1$, is the continuity solution. The second one, with $r=2$, is the parallel strong sonic shock solution for Stage 1, already studied in \cite{BretJPP2018}.
  \item The second branch pertains to $\Lambda=0$. We can extract the value of $r$ from $\Lambda=0$, namely,
  \begin{equation}\label{eq:rS1}
   r=\frac{2 X_2^2}{\sigma [3 (X_2^2-2) X_2^2+4]-6 X_2^4+10 X_2^2-4},
  \end{equation}
   and substitute in (\ref{eq:62}). This eventually gives a polynomial equation for $X_2$ only, which reads,
  \begin{equation}\label{eq:Q}
   Q(X_2) = \sum_{k=0}^4 b_k X_2^{2k},
  \end{equation}
   with,
   \begin{eqnarray}
     b_0 &=& 32 \sigma^2-64 \sigma+32, \nonumber\\
     b_1 &=& -80 \sigma^2+200 \sigma-120, \nonumber\\
     b_2 &=& 76 \sigma^2-224 \sigma+160, \nonumber\\
     b_3 &=& -30 \sigma^2+100 \sigma-84, \nonumber\\
     b_4 &=&  3 \sigma^2-12 \sigma +12.
   \end{eqnarray}
   It can be solved numerically and gives the 2 values of $\theta_2(\sigma)=\arcsin X_2(\sigma)$ plotted on Figure \ref{fig:theta2}-left. Solutions exist only for $\sigma \in [0.432, 1]$. Then Eq. (\ref{eq:rS1}) allows to compute the density jump $r$ for each $\theta_2$-branch, and plot them on Figure \ref{fig:theta2}-right.
\end{itemize}

\section{Calculation of $\beta_{\parallel 2}$}\label{ap:beta}
We need to evaluate,
\begin{equation}
\beta_{\parallel 2} = \frac{n_2  k_B T_{2\parallel}}{B_2^2/(8\pi)}.
\end{equation}
Eq. (\ref{eq:2}) gives $B_2 =B_1/\cos \theta_2$. Also, Eq. (\ref{eq:TS1T10}) gives $T_{2\parallel} =  T_e \cos^2\theta_2$. Finally, Eq. (\ref{eq:Td}) gives $T_e$. Expressing the result in terms of the dimensionless variables $r$ and $\sigma$ yields Eq. (\ref{eq:beta2}) for $\beta_{\parallel 2}$.

%\bibliographystyle{jpp}
%\bibliography{BibBret}

\end{document}